\documentclass[aps,prb,twocolumn,floats,showpacs,superscriptaddress]{revtex4-1}
\usepackage{graphicx,epsfig}
\usepackage{times}
\usepackage{graphics,dcolumn,bm,float}
\usepackage{amssymb,amsmath,rotate,color}
\usepackage[title,titletoc,toc]{appendix}
\usepackage[pagebackref=false,colorlinks,linkcolor=blue,citecolor=blue,urlcolor=magenta]{hyperref}
\DeclareMathOperator\atanh{arctanh}
\DeclareMathOperator\asinh{arcsinh}

\begin{document}
\unitlength 1 cm
\newcommand{\be}{\begin{equation}}
\newcommand{\ee}{\end{equation}}
\newcommand{\bearr}{\begin{eqnarray}}
\newcommand{\eearr}{\end{eqnarray}}
\newcommand{\nn}{\nonumber}
\newcommand{\la}{\langle}
\newcommand{\ra}{\rangle}
\newcommand{\cd}{c^\dagger}
\newcommand{\vd}{v^\dagger}
\newcommand{\ad}{a^\dagger}
\newcommand{\bd}{b^\dagger}
\newcommand{\tk}{{\tilde{k}}}
\newcommand{\tp}{{\tilde{p}}}
\newcommand{\tq}{{\tilde{q}}}
\newcommand{\eps}{\varepsilon}
\newcommand{\vk}{\vec k}
\newcommand{\vp}{\vec p}
\newcommand{\vq}{\vec q}
\newcommand{\vkp}{\vec {k'}}
\newcommand{\vpp}{\vec {p'}}
\newcommand{\vqp}{\vec {q'}}
\newcommand{\bk}{{\bf k}}
\newcommand{\bp}{{\bf p}}
\newcommand{\bq}{{\bf q}}
\newcommand{\br}{{\bf r}}
\newcommand{\bR}{{\bf R}}
\newcommand{\up}{\uparrow}
\newcommand{\dn}{\downarrow}
\newcommand{\fns}{\footnotesize}
\newcommand{\ns}{\normalsize}
\newcommand{\cdag}{c^{\dagger}}

\newcommand{\bal}{\begin{align}}
\newcommand{\eal}{\end{align}}

\newcommand{\sx}{\sigma^x}
\newcommand{\sy}{\sigma^y}
\newcommand{\sz}{\sigma^z}
\newcommand{\ket}[1]{|{#1}\rangle}
\newcommand{\mel}[3]{\langle #1|{#2}|#3\rangle}

\title{Exact phase diagram and topological phase transitions of the XYZ spin chain}

\author{S. A. Jafari}
\email{jafari@physics.sharif.edu}
\affiliation{Department of Physics, Sharif University of Technology, Tehran 11155-9161, Iran}
\affiliation{Center of excellence for Complex Systems and Condensed Matter (CSCM), Sharif University of Technology, Tehran 1458889694, Iran}
\affiliation{Theoretische Physik, Universit\"at Duisburg-Essen, 47048 Duisburg, Germany}

\begin{abstract}
Within the block spin renormalization group we are able to construct the {\em exact} phase diagram 
of the XYZ spin chain. First we identify the Ising order along $\hat x$ or $\hat y$ 
as attractive renormalization group
fixed points of the Kitaev chain. Then in a global phase space composed of the anisotropy $\lambda$ 
of the XY interaction and the coupling $\Delta$ of the $\Delta\sz\sz$ interaction
we find that the above fixed points remain attractive in the two dimesional parameter space. 
We therefore classify  
the gapped phases of the XYZ spin chain as: (1) either attracted to the Ising 
limit of the Kitaev-chain which in turn is characterized by winding number $\pm 1$ depending whether 
the Ising order parameter is along $\hat x$ or $\hat y$ directions; or (2) attracted to 
the Mott phases of the underlying Jordan-Wigner fermions which is characterized by zero winding
number. We therefore establish that the exact phase boundaries of the XYZ model in Baxter's solution
indeed correspond to topological phase transitions. The topological nature of the phase
transitions of the XYZ model justifies why our analytical solution of the three-site problem
which is at the core of the renormalization group treatment is able to produce the {\em exact}
phase diagram of Baxter's solution. We argue that the distribution of the winding numbers between
the three Ising phases is a matter of choice of the coordinate system, and therefore the Mott-Ising
phase is entitled to host apprpriate form of zero modes. We further observe that
the renormalization group flow can be cast into a geometric progression of a properly identified
parameter. We show that this new parameter is actually the size of the (Majorana) zero modes. 
\end{abstract}
\pacs{
   75.10.Pq,	
   03.65.Vf,	
   64.40.ae	
}
\maketitle
\section{Introduction}
The XYZ spin chain is the most anisotropic from of the Heisenberg spin chain 
where the coupling between $x$, $y$ and $z$ components of adjacent spins are 
generically different,
\begin{align}
\label{xyz.eqn}
   &H_{\rm XYZ} = \sum_j (J+\lambda) \sx_j \sx_{j+1} + \sum_j (J-\lambda) \sy_j \sy_{j+1}\nn\\
   &+\Delta\sum_{j}\sz_j\sz_{j+1}.
\end{align}
Baxter was able to solve the eight-vertex model~\cite{BaxterPRL,Baxter8V} from which also
the exact solution of the XYZ model follows~\cite{BaxterXYZ,McCoy}. Recent progress in the 
off-diagonal Bethe ansatz has also enabled exact solutions for arbitrary boundary field~\cite{ODBA,odba-npb}. 
The limit $\Delta=0$ is exactly solvable in terms of Jordan-Wigner (JW) fermions~\cite{LSM} where the
coupling $J$ translates into the hopping amplitude of the JW fermions, and
the coupling $\lambda$, the deviation from isotropic limit induces p-wave superconducting
pairing between the resulting spin-less JW fermions~\cite{Mattis2006}. In this limit this model and even 
generalizations of this model~\cite{NXY,Raghu2011,Sen} can be solved exactly where the non-trivial topology is
encoded in a non-zero winding number $n_w$ (of the ensuing Anderson pseudo-vector) 
manifests itself as Majorana zero modes localized at the chain ends which is best pictured in terms 
of the Kitaev chain~\cite{Kitaev2001}. 

When written in terms of the JW fermions which are particle excitations of the XY limit~\cite{LSM},
the coupling $\Delta$ will correspond to density interaction between the fermions
and hence introduces further many-body effects into the problem~\cite{Fradkin}. 
Luther noticed that the above JW mapping translates the XYZ model into the lattice version 
of the massive Thirring model~\cite{Luther}. This enabled a Bethe ansatz solution for the
massive Thirring model~\cite{Bergknoff}. 
The equivalence between massive Thirring model and
the sine-Gordon (bosonic) theory has its own rich literature~\cite{Fradkin,Tsvelik,Gogolin}. 

In the $\lambda=0$ limit we are dealing with a liquid of JW fermions interacting through $\Delta$ term
which corresponds to massless Thirring model. In this limit if the 
coupling $\Delta$ is below a certain critical value $\Delta_c$ the system remains gapless, 
but if it is stronger than $\Delta_c$ the system enters the Mott insulating phase and becomes gapped.
The picture in the $\lambda=0$ is therefore that of a critical line that ends at a 
a Berezinskii-Kosterlitz-Thouless (BKT) 
point $\Delta_c$ and the algebraic correlations of the gapless phase are rendered exponential with a
correlation length determined by the spectral gap.
The field theory value of $\Delta_c$ is $\pi J/2$ while the exact solution gives $\Delta_c=J$~\cite{Fradkin}. 
The XXZ limit that would correspond to massless Thirring model has been analyzed from
spin systems and entanglement points of view~\cite{Langari2009,Song2013} where
the critical value is obtained to be $\Delta_c=J$. 
In the $\lambda=0$ situation a Dzyaloshinskii-Moriya interaction of strength $D$ can be added 
to the above $XXZ$ form which results in a gapless line separating the spin-fluid phase from
ferromagnetic and/or anti-ferromagnetic Ising phases depending on the sign of $\Delta$~\cite{Langari2009}.
The XXZ model in external magnetic field was also found to posses a critical line separating the 
saturated magnetized phase from the Ising phase at large $\Delta$~\cite{Langari98}.

When both the pairing gap $\lambda$ and the interaction parameter $\Delta$ compete with each other, 
in the limit where $\Delta$ dominates we expect a gapped phase that corresponds to the
Mott insulating phase of the corresponding Thirring model. When $\Delta$ is negligible the parameter $\lambda$ being
a pairing strength of the JW fermions, gives rise to a (p-wave) pairing gap. 
The above gapped phases must be separated by a 
gapless line in the plane of $(\Delta, \lambda)$~\cite{denNijs}. Indeed Ercolessi and coworkers using the exact
solution of Baxter~\cite{BaxterPRL,Baxter8V,BaxterXYZ} calculated the Renyi entropy and identify lines
of essential singularity~\cite{Ercolessi2011} ending at tricritical points where the lines join.
They find that two of the tricritical points are conformally invariant~\cite{Ercolessi2011,Ercolessi2013}. 

Earlier attempt using block-spin renormalization group (RG) was undertaken by Langari~\cite{Langari2004} who
used a two site cluster to study the phase diagram of the model in presence of a transverse field. 
However clusters with even number of sites are not able to provide a true Kramers doublet degenerate 
ground states. Therefore the two low-energy states used in the above work consists in, one singlet, and one
triplet (split by the Zeeman coupling) which are not connected by the time reversal operation. 
In this work, by introducing a conserved charge and making use of the mirror symmetry we are able
to break the Hilbert space of the three site problem into blocks of maximum three dimension which 
can then be analytically diagonalized. Such an analytical solution of the three site cluster of the XYZ 
model enables us to construct a phase portrait of the XYZ model when both $\Delta$ and $\lambda$ are present. 
Before discussing the result let us note that in the Ising limit, ferromagnetic (FM) and antiferromagnetic (AFM) 
Ising chains are related by a simple unitary transformation at alternating sites. We therefore colloquially use
the term "Ising order" (IO) to refer to both magnetization of the FM Ising case and the staggered magnetization 
in the AFM Ising case. Now let us summarize the outcome of our block spin RG: 
(1) First within the XY model (i.e. $\Delta=0$ case) we are able to identify Ising limits
corresponding to IO along $\hat x$ or $\hat y$ as two attractive RG fixed points of 
the Kitaev chain. This therefore attaches topological significance to IO along $\hat x$ 
or $\hat y$ which are characterized with a winding number $\pm 1$, respectively.
(2) When we turn on the coupling $\Delta$ for a region that $\Delta$ dominates again we have
an attractive Ising fixed point for very large $\Delta$ which however is characterized
with a zero winding number.
To emphasize the non-zero winding number, we call states with IO along $\hat x$ and $\hat y$ the
Kitaev-Ising (KI) fixed points, while the sate with IO along $\hat z$ direction the Mott-Ising (MI) 
fixed point.
(3) The gapless lines that separate regions with winding numbers of $0,\pm 1$ within our
block spin RG treatment using three-site cluster coincides with the {\em exact} phase 
boundaries obtained from the Baxter's exact solution. 

The main message of this paper will be that the XYZ model has essentially three
different gapped phases characterized with winding numbers $n_w=0,\pm 1$. 
The phase portrait of the model can be described by BKT repellers and Ising attractors 
with different topological charge. 
In the JW representation the $n_w=0$ phase corresponds to a Mott insulating phases,
while the other two correspond to p-wave superconducting states. While the Mott phase 
of JW fermions corresponds to IO order along $\hat z$, the topologically non-trivial
superconducting phase will correspond to IO along $\hat x$ or $\hat y$. 
The winding number of the gapped phases changes between the above three values upon 
crossing the critical (gapless) lines. The essential significance of the topology is that
the topological charges
are not sensitive to many details, including the size of the cluster as long as it does
not miss the essential symmetries of the Hamiltonian. This explains why a three-site problem
that correctly embeds the Kramers doublet structure of the ground states is capable of 
capturing the exact phase diagram of the model. 

\section{formulation}
Let us start by stating simple, but very important property of the XYZ model.
For the XYZ Hamiltonian the quantity $\zeta=\prod_{i=1}^N \sz_i$
is a constant of motion. This is straightforward to see: Assume any arbitrary state
with some arrangements of $\up$ and $\dn$ spins. Operating with the XYZ 
Hamiltonian on it since there are two consecutive $\sx$ or two consecutive $\sy$
operations on the spins of the system, the total number of spin flips is even and hence
either two $\up$ are turned into two $\dn$ (or vice versa) or the $\up\dn$ is
turned into $\dn\up$ which does not change the value of $\zeta$. 
This observation indeed will allow us to analytically nail down the three-site problem and
write down its ground state properties in the closed form. 
Two possible $\zeta=\pm 1$ values correspond to number parity of JW fermions. 
Indeed the JW transformation~\cite{Mattis2006},
\be
   \sz_j=1-2\cdag_jc_j,~~\sx_j=e^{i\phi_j} (c_j+\cdag_j),~~\sy_j=-i e^{i\phi_j} (c_j-\cdag_j),
\ee
where $\phi_j$ is the phase string defined as $\phi_j=\pi \sum_{i < j} \cdag_i c_i$
converts the above Hamiltonian to,
\begin{align}
   H=&2\sum_{j} (J \cdag_j c_{j+1}+\lambda c_jc_{j+1}+{\rm h.c.})\nn\\
   &+\Delta\sum_j(2n_j-1)(2n_{j+1}-1)
   \label{JW.eqn}
\end{align}
The XY part of the above Hamiltonian ($\Delta=0$) when rewritten in terms of the 
following Majorana fermions $a_j=c_j+c^\dagger_j$ and $b_j=i(c_j-c^\dagger_j)$ becomes,
\be
   H=i\sum_j(J+\lambda)a_jb_{j+1}+(\lambda-J)b_ja_{j+1}.
   \label{mfH.eqn}
\ee
In the Ising limit $J=+(-)\lambda$, the above Hamiltonian couples every Majorana fermion (MF) $a$ with
a Majorana fermion $b$ to its right (left), leaving a $b$ MF at the left (right)
of the chain, and one $a$ MF at the right (left) of the chain~\cite{Kitaev2001,NXY,Alicea},
as depicted in the inset of Fig.~\ref{vflow.fig} (See also discussion following Eq.~\eqref{diffv.eqn}).

\subsection{Three site problem}
Consider three sites labeled by $j=0,1,2$ for which we would like to construct the
matrix representation of the Hamiltonian~\eqref{xyz.eqn} in the $\sz$ basis which is
a $8$ dimensional space. Conservation of $\zeta$ breaks the Hilbert space into two $4$ dimensional blocks.
The $\up$ spin configuration corresponds to $\sigma^z=+1$ and hence $c^\dagger c=0$. 
In the $\zeta=+1$ sector, there are four states, $\ket{\up\up\up}$, $\ket{\up\dn\up}$,
$\ket{\up\up\dn}$ and $\ket{\dn\up\up}$.
The first two are even with respect to reflection with respect to the middle site.
Therefore the last two better be combined into even and odd combinations to give
the following symmetry adopted basis in $\zeta=+1$ sector:
\begin{align}
&\ket1^+=\ket{\up\up\up},~~\ket2^+=\ket{\up\dn\up},\nn\\
&\ket3^+=\left(\ket{\up\up\dn}+\ket{\dn\up\up}\right)/\sqrt2,\nn\\
&\ket4^+=\left(\ket{\up\up\dn}-\ket{\dn\up\up}\right)/\sqrt2.\nn
\end{align}
Similarly for the $\zeta=-1$ sector all we need is to replace $\up$ and $\dn$ spins. 
The Hamiltonian being invariant under reflection with respect to the middle site does not mix
even and odd-parity states and hence $\ket4^+$ is already an eigen-state.
Straightforward application of the multiplication table~\ref{mult.tab} to
all bonds on the above state reveals the energy of $\ket{4}^+$ state to be zero. 
\begin{table} [ht]
\centering
\begin{tabular}{|c|c|c|c|c|c|}
\hline
$\ket{{s_1s_2}}$	&$\sx_1\sx_2$	&$\sy_1\sy_2$	&$\sz_1\sz_2$	&$\sx_1\sy_2-\sy_1\sx_2$ &$\sz_1\sz_2$\\
\hline 
$\up\up$	&$\dn\dn$	&$-\dn\dn$	&$+\up\up$	&$0$		&$+\up\up$\\
$\dn\dn$	&$\up\up$	&$-\up\up$	&$+\dn\dn$	&$0$		&$+\dn\dn$\\
$\up\dn$	&$\dn\up$	&$+\dn\up$	&$-\up\dn$	&$-2i\dn\up$	&$-\up\dn$\\
$\dn\up$	&$\up\dn$	&$+\up\dn$	&$-\dn\up$	&$+2i\up\dn$	&$-\dn\up$\\
\hline
\end{tabular}
\caption{Multiplication table summarizing the effect of various operators on two-spin states.}
\label{mult.tab}
\end{table}
Using this table we can operate with the Hamiltonian in the
three dimensional space of even states which (for both $\zeta=\pm1$ sectors) gives,
\be
H=2
\begin{bmatrix}
   \Delta		&0		&\sqrt2\lambda	\\
   0			&-\Delta	&\sqrt2 J	\\
   \sqrt2\lambda	&\sqrt2 J	&0		
\end{bmatrix}.
\label{3site.eqn}
\ee
The above form is suggestive as it corresponds to two levels at $\pm2\Delta$
coupled by "hybridization" of strengths $2\sqrt\lambda$ and $2\sqrt J$ to a third level 
at energy $0$. The observation at the three-site level is that the transformation 
$\Delta\to -\Delta$ accompanied by $\lambda\leftrightarrow J$ does not change the spectrum.
However this simple observation helps us to identify a symmetry of the XYZ
spin chain in general of any size:
 Indeed in Eq.~\eqref{xyz.eqn} this comes from the unitary 
transformation $\sx_j\to\sx_j$, $\sy_j\to(-1)^j\sy_j$, $\sz_j\to (-)^j\sz_j$ that preserves the $SU(2)$
algebra. The above transformation when translated into the language of JW fermions
is actually a particle-hole transformation {\em only in one sub-lattice} which induces the
change in the role of $\lambda$ and $J$. 

\subsection{Block spin renormalization group transformations}
When we are dealing with odd number of sites the ground state will belong
to a Kramers doublet. In the language of JW fermions this correspond to odd number-parity
of JW fermions which has a chance to produce Majorana fermions. In the case of XYZ
model where the quantity $\zeta$ is conserved, the two degenerate
ground states correspond to $\zeta=\pm 1$.
The basic idea of block-spin RG is to construct the matrix elements of 
the operator ${\cal O}=\sigma^x_j,\sigma^y_j,\sigma^z_j$ in the space
of the these doublet $\{\ket{\phi_\zeta}\}$, $\zeta=\pm$ as
\be
   \begin{bmatrix}
   \mel{\phi_+}{\cal O}{\phi_+} 	&\mel{\phi_+}{\cal O}{\phi_-} \\
   \mel{\phi_-}{\cal O}{\phi_+} 	&\mel{\phi_-}{\cal O}{\phi_-}\\
   \end{bmatrix}
\ee
which being $2\times 2$ matrix can again be rewritten in terms of Pauli matrices 
$\sigma'^a,~a=x,y,z$.  This can be interpreted as new spin-half degrees of 
freedom on a coarse-grained lattice~\cite{Langari98}. The relation between the
new couplings and the old couplings is the block-spin RG transformation. 
In the following two sections we proceed with implementation of this RG program
first for the limiting cases from which we learn about topology of the attractors,
and next for a general case. 

\section{Limiting cases: XY and XXZ chains}
Before considering the solution of the problem in the most general case $\lambda\ne 0, \Delta\ne 0$,
it is instructive to consider the special cases first to establish the Ising limit 
of the Kitaev-chain Hamiltonian as a renormalization group fixed point. 
We provide analytic solutions for the RG flow and will identify a length scale associated
with the MFs.
Then we will proceed to construct the whole picture for the phase diagram.

\subsection{The XY limit: $\lambda\ne 0$ and $\Delta=0$}
\label{XY.sec}
This limit is exactly solvable by the JW transformation giving a half-filled
chain of JW fermions hopping with amplitude $J$ and with p-wave superconducting pairing
of strength $\lambda$ between them~\cite{Kitaev2001}. This superconductor belongs to BDI~\cite{NXY}  
class of topological superconductors characterized by a winding number. Changing the sign of
$\lambda$ amounts to changing the winding number. For any non-zero $\lambda$ we have a 
topologically non-trivial gapped (superconducting) phase. The gapped phases
corresponding to positive and negative values of $\lambda$ are separated by a gap closing at
$\lambda=0$. To set the stages for the general case of the XYZ spin chain, first of all, 
let us see how can the above picture be reproduced in the block spin RG language.

In the XY limit where $\Delta=0$, the eigenvalue equation~\eqref{eigenval.eqn} gives
three eigenvalues,
\be
   \omega_m=m2\sqrt 2 \sqrt{J^2+\lambda^2}, ~~~~m=0,\pm 1,\nn
\ee
with corresponding eigen-states in $\zeta=\pm 1$ sectors,
\begin{align}
   &\ket{\psi_0}^\zeta=\frac{1}{\sqrt{J^2+\lambda^2}}
   \left(-\lambda\ket{1}^\zeta+J\ket{2}^\zeta\right),\nn\\
   &\ket{\psi_\pm}^\zeta=\frac{1}{\sqrt2\sqrt{J^2+\lambda^2}}
   \left(J\ket{1}^\zeta+\lambda\ket{2}^\zeta\right)+\frac{\pm1}{\sqrt2}\ket{3}^\zeta.\nn
\end{align}
 Obviously the Kramers doublet of ground states is given by 
$\ket{\phi_\zeta}=\ket{\psi_-}^\zeta$
for $\zeta=\pm 1$, which explicitly reads,
\bearr
   \ket{\phi_+}&&=\bar a\ket{\up\up\up}+\bar b\ket{\dn\up\dn}-0.5\ket{\up\dn\dn}-0.5\ket{\dn\dn\up}\label{kdp.eqn}\\
   \ket{\phi_-}&&=\bar a\ket{\dn\dn\dn}+\bar b\ket{\up\dn\up}-0.5\ket{\dn\up\up}-0.5\ket{\up\up\dn}\label{kdm.eqn}\\
   \bar a&&=\frac{\lambda}{\sqrt{2(J^2+\lambda^2)}},~~~\bar b=\frac{J}{\sqrt{2(J^2+\lambda^2)}}.
   \label{ab.eqn}
\eearr
Evaluating the matrix elements of the original spin variables linking the adjacent blocks gives,
\bearr
   &&\sigma^x_{0,2} =(-\bar a-\bar b)\sigma'^x\\
   &&\sigma^y_{0,2} =(-\bar a+\bar b)\sigma'^y
\eearr
where we only need the sites $0,2$ at the boundary of three site cluster.
The $\vec\sigma'$ denotes the coarse-grained Pauli matrices for a given block which describe
the fluctuations in the ground state of Kramers doublets $\ket{\phi_\zeta}$. These coarse-grained
Pauli matrices are mutually connected to neighboring
blocks only through sites $0,2$. Note that it is not surprising that the spin variables at sites
$0,2$ of the cluster transform identically under coarse-graining, as the three site cluster has a reflection
symmetry with respect to the middle site which has been nicely manifested in the Kramers doublet ground 
states~\eqref{kdp.eqn} and~\eqref{kdm.eqn}. Therefore the interaction terms connecting neighboring blocks are
transformed as
\bearr
   && (J+\lambda)\sigma^x_0\sigma^x_2 \to (J+\lambda)(-\bar a-\bar b)^2\sigma'^x_n\sigma'^x_{n+1}\\
   && (J-\lambda)\sigma^y_0\sigma^y_2 \to (J-\lambda)(-\bar a+\bar b)^2\sigma'^y_n\sigma'^y_{n+1}
\eearr
Using Eq.~\eqref{ab.eqn} allows to identify the coarse-grained couplings $J'$ and $\lambda'$ as,
\be
   J'\pm\lambda'=-\frac{(J\pm\lambda)^3}{2(J^2+\lambda^2)}
\ee
or equivalently,
\be
    J'=\frac{J}{2}\frac{J^2+3\lambda^2}{J^2+\lambda^2},~~~~~
    \lambda'=\frac{\lambda}{2}\frac{\lambda^2+3J^2}{\lambda^2+J^2}.
    \label{Jlambda.eqn}
\ee
These equations are invariant under $J\leftrightarrow \lambda$
which is actually due to symmetry operation discussed under Eq.~\eqref{3site.eqn}. 
Since the physics of XY Hamiltonian is given only by the ratio of the above parameters, 
dividing the above flow equations gives,
\be
   y'=y\frac{3+y^2}{1+3y^2},~~~~~~y=\frac{\lambda}{J}
   \label{recy.eqn}
\ee
Note that the above flow equation is invariant under $y\to y^{-1}$. 
This is natural since the above transformation is the same as $\lambda\leftrightarrow J$.
Obviously $y^*=0$ is
a fixed point, which then due to this symmetry implies that it is equivalent to $y^*=\infty$ fixed point. 
The invariance of fixed points under $y^* \to 1/y^*$ implies that there could be fixed points
at $y^*=\pm 1$ as well. This can be explicitly obtained by constructing the difference equation~\cite{Strogatz}
for the variable $y$ that reads,
\be
   y_{n+1}-y_n=2y_n\frac{1-y_n^2}{1+3y_n^2}.
   \label{diffv.eqn}
\ee
\begin{figure}[th]
\includegraphics[width=0.45\textwidth]{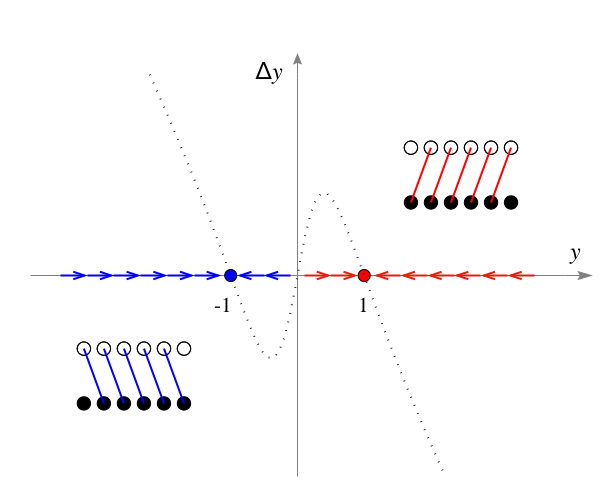}
\caption{(Color online) Fixed points of the Kitaev (XY) chain:
Phase portrait of the flow equation~\eqref{recy.eqn} is characterized by 
repulsive fixed points at $0$ and $\pm\infty$ and two attractive
fixed points at $y^*=\pm 1$ that are related by $\lambda\to -\lambda$ transformation. 
The direction of arrows correspond to the sign of the right hand side of the difference 
equation~\eqref{diffv.eqn}. The two attractive fixed points at $y^*=+1$ (red) corresponds to
$J_\infty=\lambda_\infty$ while $y^*=-1$ (blue) corresponds to $J_\infty=-\lambda_\infty$.
The arrows represent that way MF of type $a$ is combined with a neighboring MF of type $b$
which then leaves a pair of spatially separated MF end modes. 
The Kitaev-Ising fixed point denoted by red (blue) correspond to IO
along $\hat x$ ($\hat y$) direction. 
}
\label{vflow.fig}
\end{figure}
The zeros of the right hand side give the fixed points~\cite{Strogatz} $y^*=\pm 1$ which are both 
attractive fixed points along with a repulsive fixed point at $y^*=0$. 
The invariance of the flow equations under $y\to 1/y$ 
imply that the infinity is also repulsive fixed point.  
Under this symmetry operation, the attractive fixed points at $y^*=\pm 1$
map to themselves. This has been plotted in Fig.~\ref{vflow.fig}. 

The interpretation of this flow diagram is as follows: first of all, $\lambda=0$ corresponding to gapless
phase of the filled Fermi sea of JW fermions is a repulsive (unstable) fixed point. 
In the language of the JW fermions 
this is nothing but the instability of a Fermi system with respect to the superconducting
pairing interactions (i.e. $\lambda$). 
Any non-zero $\lambda$ flows ultimately to the $\pm J$ depending on the sign of $\lambda$. 
In terms of $J_{x,y}=J\pm\lambda$ it means that the smallest of $J_x$, $J_y$ is renormalized
to zero, and the fixed point is an Ising chain polarized along $\hat x$ or $\hat y$ direction.
In such an Ising ground state, every MF of a given
type is paired with a MF of opposite type in either right (red, fixed point $\lambda/J=+1$)
or left (blue, fixed point $\lambda/J=-1$). To clarify 
this let us repeat the analysis of Kitaev~\cite{Kitaev2001}:
For the fixed point at $y^*=1$ denoted by red in Fig.~\ref{vflow.fig} the Majorana representation of the
Hamiltonian is,
$$
  H_{{\rm FP},y^*=+1}=4iJ_{\infty}\sum_j a_jb_{j+1}.
$$
In terms of new fermions denoted by red links in the inset of Fig.~\ref{vflow.fig} , i.e.
$f^\dagger_j=a_j+ib_{j+1}$ the above Hamiltonian is simply $2J\sum_{j=0}^{N-1}f^\dagger_jf_j$
which does not include the $a_0$ nor $b_{N-1}$ which are denoted as unpaired circles in
the inset of Fig.~\ref{vflow.fig}. Similarly the other fixed point at $y^*=-1$ corresponds 
to a Kitaev chain where every MF of type $a$ is paired with the $b$ MF to its left,
leaving again two unpaired MFs at the chain end~\cite{Akbari}. The two fixed points $y^*=\pm 1$ are related
by the $\lambda\to -\lambda$ transformation which in the language of original spins
amounts to a rotation around $\hat z$ axis of the spins, $\sx\to\sy$ and $\sy\to -\sx$.

Therefore the KI fixed points at $y^*=\pm 1$ with their non-trivial topology
that ultimately spawn two sharply localized Majorana zero modes at the two ends of the chain
are the fixed points of the XY (Kitaev) chain. This interpretation can be understood intuitively:
For e.g. the $y=y^*=1$ point the resulting MFs are sharply (Dirac delta) localized at the
two ends of the chain. If one solves the zero mode eigenvalue problem with a simple Z-transform
method~\cite{NXY} one finds that away from the fixed point, one still has the MFs at the
two ends, but this time they are exponentially decaying instead of being Dirac-delta localized 
(see Eq.\ref{recpsi.eqn}). 
Take any chain with $y$ away from $y^*$ with its exponentially localized pair of MFs at the
chain ends, and look at the MFs at larger length scales: After every scale transformation
the MF wave function will be more and more localized. This means that at very large length
scales the MF will look like a Dirac-delta localized MF. This is the meaning of flowing
towards Dirac-delta-localization -- i.e. the $y=y^*$ -- point. 

Let us now show that the recursive relation~\eqref{recy.eqn} can be explicitly solved.
Let us change of variable $y$ as,
\be
   y_n=\tanh u_n=\frac{e^{u_n}-e^{-u_n}}{e^{u_n}+e^{-u_n}}
   \label{utrans.eqn}
\ee
which after a little algebra renders the recursive Eq.~\eqref{recy.eqn} to
a simple geometric progression for the new variable,
\be
   y_{n+1}\equiv \tanh(u_{n+1})=\frac{e^{3u_n}-e^{-3u_n}}{e^{3u_n}+e^{-3u_n}}=\tanh (3u_{n}),
\ee
whose solution $u_n=3^nu_0$ implies,
\be
   y_n=\tanh\left(3^n\atanh(y_0)\right),
\ee
where the system size at the $n$'th RG level, $\ell_n=3^n$ appears quite naturally in this solution. 

The relation $u_n=\ell_n u_0$ suggests that $u$ must be some sort of length scale. 
The question is, what kind of length scale is it? The answer to this question is surprisingly 
simple and physical: Assume that Eq.~\eqref{mfH.eqn} has a zero mode solution of 
the form which has amplitude $\psi^{(b)}_j$ at every site $j$ of the lattice. Then it
will satisfy the recursive relation,
\be
   (1+y)\psi^{(b)}_{j+1}+(1-y)\psi^{(b)}_{j-1}=0.
   \label{recpsi.eqn}
\ee
This equation is solved by the exponentially decaying ansatz,
$\psi_{j}^{(b)}=\exp(-j/u)$, where,
\be
   u=\frac{1}{2}\ln\left(\frac{y+1}{y-1}\right).
   \label{mflength.eqn}
\ee
This is nothing but the transformation~\eqref{utrans.eqn} in disguise.
{\em This relation enables us to interpret the quantity $u$ as the length scale
associated with the zero modes.}

This solution enables us to figure out the flow of the energy gap per site ($u_n=\ell_n u_0$),
\be
   E_{g,n}=\frac{2\sqrt2}{3} J_n\sqrt{1+y_n^2}=\frac{2\sqrt2}{3} J_n\frac{\sqrt{\cosh(2u_n)}}{\cosh(u_n)}
   \label{gapflow.eqn}
\ee
The symmetry of the problem under $\lambda\to -\lambda$
is reflected at this stage in a symmetric functional dependence on $u_0$. 
For $y_0=\tanh u_0=0$ the above equation gives $E_{g,n}=2\sqrt2 J_n$. On the other hand
the flow equation~\eqref{Jlambda.eqn} for $J$ will become $J_{n+1}=J_n/2$
which gives $J_n=J_0(1/2)^n$ vanishing for $n\to\infty$ and hence giving a gapless system, in 
agreement with the exact solution of the XY model~\cite{NXY}. 
For any non-zero value of 
$y_0$, with the behavior of gap for $u_0\sim y_0\to 0$ in mind, the ratio of
$\cosh$ terms in the above expression for large enough length scales is always
close to $1$ and therefore the essential factor that determines the
behavior of gap is the behavior of $J_n$ at large length scales.
Therefore we need to solve the recursive relation~\eqref{Jlambda.eqn} for $J$ which is,
\be
   J_{n+1}=J_n\left(1-\frac{1}{e^{2u_n}+e^{-2u_n}}\right),
\ee
where $u_n$ is a geometric progression corresponding to the size of MF. 
The solution to the above equation is,
\begin{align}
   J_n&=J_0\prod_{i=1}^n \left(1-\frac{1}{e^{2u_i}+e^{-2u_i}}\right),\nn\\
   &\approx J_0 \prod_{i=1}^{n_{\rm conv}} \left(1-\frac{1}{e^{2u_i}+e^{-2u_i}}\right),
\end{align}
where we have used the fact that
due to geometric progression nature of $u_n$, the $e^{-u_n}$ rapidly converges to zero
in $n_{\rm conv}\sim -(\ln u_0/\ln 3)$ steps. 
This causes the terms in parenthesis to come close to $1$ which prevents
vanishing of the $J$ in large length scales. 
A lower bound for the $J$ in large $n$ limit is obtained for very small $u_0$
by setting $u_0=0$ in all the $n_{\rm conv}$ terms in the parenthesis which 
gives the lower bound for $J$ and hence $E_g$ at large length scales as,
\be
   E_g>\left(\frac{1}{2}\right)^{n_{\rm conv}}=u_0^{\ln 2/\ln 3}\sim y_0^{0.63093}
   \label{gapexponentest.eqn}
\ee

\begin{figure}[h]
\includegraphics[width=0.49\textwidth]{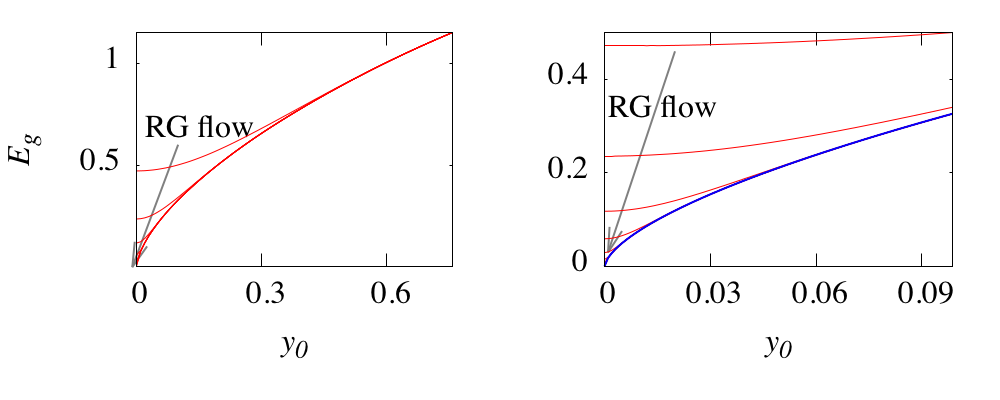}
\vspace{-0.4cm}
\caption{(Color online) Flow of the energy gap (per site) as a function of $y_0$. 
The red curve is the fit to the numerical data obtained from iteration of Eq.~\eqref{gapflow.eqn}
which gives $1.4076 y_0^{0.631}$.
}
\label{xygap.fig}
\end{figure}
Numerical evaluation of the iterative equation quickly converges and
results in plots represented in Fig.~\ref{xygap.fig}. The trends of the
gap curves as function of the initial gap parameter $y_0$ are indicated by 
arrows in both panels. In the right panel we have magnified the range of
$y_0<0.1$ to which the limiting (blue) curve given by $E_g^{\rm fit}=1.4076 y_0^{0.631}$
gives a perfect fit. Indeed zooming in by a further order of magnitude does
not change the exponent up to third decimal point, which indicates the quality of fit. 
It is remarkable that the above exponent is so close to the lower bond estimated
in Eq.~\eqref{gapexponentest.eqn}. 
The exact solution from the JW transformation gives a superconducting state 
with pairing potential proportional to $y_0$ and hence the exact gap exponent
is actually $1$. However, the value of $\ln 2/\ln 3$ obtained above gives 
the finite size value of the gap exponent. 

A final note is that the behavior of flow Eq.~\eqref{recy.eqn} near the critical 
point $y^*=0$ (where the gap is zero) is given by $y_{n+1}=3y_n$. Assuming a divergent correlations length
$\xi(y)\sim|y|^{-\nu}$ and demanding $3\xi(y_{n+1})=\xi(y_n)$ gives the correlation 
length exponent for the XY model by $\nu=1$.

\subsection{XXZ limit: $\lambda=0$ and $\Delta\ne 0$}
This limit indeed has been studied much in its spin, 
fermionic (massless Thirring model), and its bosonic disguise (Sine-Gordon model). 
To slightly generalize it, let us add a Dzyaloshinskii-Moriya (DM)
interaction~\cite{Langari2009} and on every link we consider the operators,
\be
   J\sigma^x_j\sigma^x_{j+1}+J\sigma^y_j\sigma^y_{j+1} +\Delta \sigma^z_j\sigma^z_{j+1}
   +D\left(\vec\sigma_j\times\vec\sigma_{j+1}\right)_z.
   \label{xxzdm.eqn}
\ee
The Hamiltonian \eqref{xxzdm.eqn} not only conserves the the fermion number parity
$\zeta=\prod_j\sz_j=\pm 1$, but also is invariant with respect to rotation around
$z$ axis, and hence the total $\sz$ is also conserved. 
To search for our Kramers doublets we are interested in a sector with total $\sz$ equal
to $\pm 1$. In the $\zeta=+1$ sector we define the following states,
\be
   \ket{1}=\ket{\dn\up\up},~~~~~\ket{2}=\ket{\up\dn\up},~~~~~\ket{3}=\ket{\up\up\dn}.
\ee
The $\zeta=-1$ space is similarly obtained by flipping every spin. With respect to the DM interaction
as can be seen from table~\ref{mult.tab}, an imaginary factor $i$ as $iD$ is involved. In this case,
changing the sign of $\zeta$ amounts to $i\to -i$, i.e. the complex conjugation. 
A formal way to see this in general is that $\zeta$ is the eigenvalue of the operator $Z=\prod_j \sz_j$.
The $\zeta$ label of a state with odd number of sites 
changes if the operator $X=\prod_j\sx_j$ acts on it. It is readily
seen that $X$ commutes with $\sum_i \sx_i\sx_{i+1}$ and  $\sum_i \sy_i\sy_{i+1}$ terms while
it anti-commutes with DM terms of  $\sum_i \sx_i\sy_{i+1}$ type. Therefore for any state $\ket{\psi}$,
there exists a state $\ket{\psi'}=X\ket{\psi}$ whose values of $D$ are negative of each other, 
and hence in the $\zeta=-1$ sector instead of $J+iD$ one has $J-iD$. More compactly for any $\zeta$, the DM
interaction upgrades $J$ to $J+i\zeta D$. 

The effect of each individual term of the above Hamiltonian on a two-spin state 
is summarized in table~\ref{mult.tab}. 
For $\zeta=1$ sector, the effect 
of the above Hamiltonian on various states can be easily seen to be,
\bearr
   && H\ket{1}=2(J+iD)\ket{2},\nn\\
   && H\ket{2}=2(J-iD)\ket{1}-2\Delta\ket{2}+2(J+iD)\ket{3},\nn\\
   && H\ket{3}=2(J-iD)\ket{2},\nn
\eearr
which gives the matrix representation,
\be
   H_{\rm 3-site}^{\rm XXZDM}=2
   \begin{bmatrix}
   0		&\xi^*		&0\\
   \xi		&-\Delta	&\xi^*\\
   0		&\xi		&0
   \end{bmatrix},
\ee
where as emphasized $\xi=J+iD\equiv r\exp{i\theta}$ combines $J,D$ into a single complex parameter.
The eigenvalues of the above matrix for the $\zeta=+1$ and $\sigma^z=1$ sector are,
\begin{align}
   &\eps_m=-|m|\Delta+m\sqrt{\Delta^2+8r^2},~~~~m=0,\pm1.
   \label{epsm.eqn}
\end{align}
For both negative and positive values of $\Delta$ the ground state corresponds to 
$m=-1$, and asymptotically approaches the first excited state at $0$ for $\Delta\to -\infty$,
but never touches it. Therefore the ground state doublet is given by,
\begin{align}
   \label{kdxxzdm.eqn}
   &\ket{\phi_+}=
   \tilde b\ket{\up\dn\up}+\tilde ce^{-i\theta}\ket{\dn\up\up}+\tilde ce^{+i\theta}\ket{\up\up\dn},\\
   \label{kdxxzdm2.eqn}
   &\ket{\phi_-}=
   \tilde b\ket{\dn\up\dn}+\tilde ce^{+i\theta}\ket{\up\dn\dn}+\tilde ce^{-i\theta}\ket{\dn\dn\up},\\
   & \tilde b=\frac{4r}{\cal N},~~~\tilde c=\frac{\eps+2\Delta}{\cal N},~~~\eps=-\Delta-\sqrt{\Delta^2+8r^2},\\
   &{\cal N}^2=4\sqrt{\Delta^2+8r^2}\left(\sqrt{\Delta^2+8r^2}-\Delta\right).
\end{align}
We have used the fact that switching between $\zeta=\pm 1$ is equivalent to complex conjugation which
replaces $e^{i\theta}$ and $e^{-i\theta}$. Let us emphasize again that as far as
the multiplication table~\ref{mult.tab} is concerned the effect of introducing the DM interaction is to replace
$J\to \xi =J+iD\equiv r\exp(i\theta)$. This looks like a global gauge transformation by an angle $\theta$ 
when expressed in terms of the JW fermions that modulates the hopping. Let us see how does it show up
in the RG language.

The matrix elements of $\vec\sigma_0$ operator in the Kramers doublet space is
summarized as,
\begin{align}
   &\sigma^{x,y}_0 =\sigma'^{x,y} 2\tilde b\tilde c \exp(-i\theta),~~~\sz_0=\sigma'^z \tilde b^2,\nn\\
   &\sigma^{x,y}_2 =\sigma'^{x,y} 2\tilde b\tilde c \exp(+i\theta),~~~\sz_2=\sigma'^z \tilde b^2.\nn
\end{align}
Note that as expected from Eq.~\eqref{kdxxzdm.eqn} and~\eqref{kdxxzdm2.eqn}, the sites $0$ and $2$ at two
ends of the three site cluster are related by $e^{-i\theta}\to e^{i\theta}$.
The above matrices give the couplings between the coarse-grained spin variables as,
\begin{align}
   J'=4J\tilde b^2\tilde c^2,~~~D'=4D\tilde b^2\tilde c^2,~~~\Delta'=\Delta \tilde b^4.
   \label{brgxxzdm1.eqn}
\end{align}
Note that the spin at site $0$ of a given cluster is connected to spin at site $2$ of the neighboring
cluster which leads to cancellation of the DM phases $\theta$, thereby making $J$ and
$D$ scale identically. The above equation in terms of the combined complex variable 
$\xi=J+iD$ becomes,
\be
   \xi'=4\xi\tilde b^2\tilde c^2,~~~\Delta'=\Delta \tilde b^4,
   \label{brgxxzdm2.eqn}
\ee
where $b$ and $c$ are real and incorporate no phase to $\xi$ through the RG scaling. 
The above equation is remarkable in that it implies that the magnitude $r=\sqrt{J^2+D^2}$
scales the the same way as $J$, but the phase $\theta$ does not change:
\be
   r'=4r \tilde b^2\tilde c^2,~~~\theta'=\theta,~~~\Delta'=\Delta \tilde b^4.
   \label{brgxxzdmr.eqn}
\ee
The fact that in the one-dimensional version of the DM interaction, 
the $\theta$ does not change as the length scale is changed is actually
a manifestation of the fact that the non-zero $\theta$ is basically a 
global gauge transformation of the $\theta=0$ limit.
As in the rest of the paper we wish to compare the results of the XXZ model 
against the XYZ, let us fix $\theta=0$ (corresponding to $D=0$). 

\begin{figure}[t]
\includegraphics[width=0.48\textwidth]{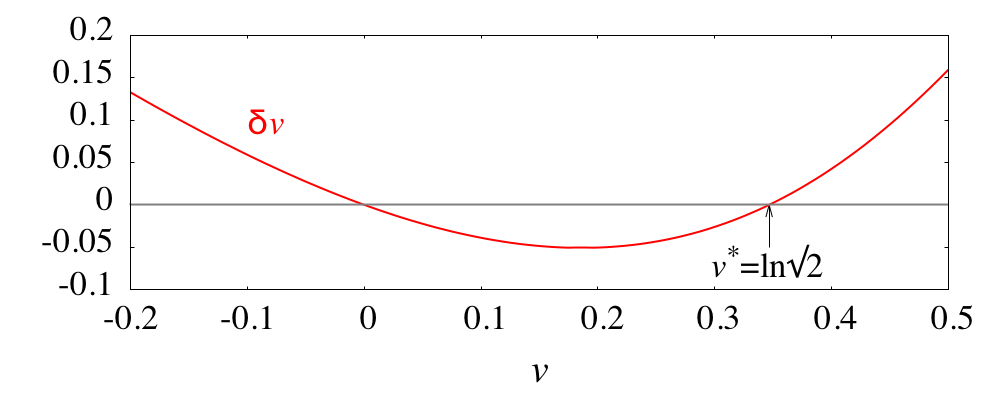}
\caption{(Color online) Fixed points of the flow equations for XXZ model.
The (non-interacting) fixed point at $v^*=0$ is attractive, while the BKT
fixed point corresponding to $v^*=\ln\sqrt 2$ is repulsive.
}
\label{xfixed.fig}
\end{figure}

In terms of $z=\frac{1}{\sqrt8}\frac{\Delta}{J}$,
the flow equations~\eqref{brgxxzdmr.eqn} becomes,
\begin{align}
   z'=\frac{z}{2}\left({\sqrt{z^2+1}+z}\right)^2,~~~~J'=\frac{J}{2}\frac{1}{1+z^2} .
   \label{rgz.eqn}
\end{align}
This naturally suggests to define,
\be
   z=\sinh v,
\ee
in terms of which the flow equations~\eqref{rgz.eqn} is simplified to,
\begin{align}
   \sinh v'=\frac{1}{2}\sinh v ~e^{2v},~~~J'=\frac{J}{2\cosh^2 v}
   \label{vflow.eqn}
\end{align}
In terms of $z$ the energy gap per site can be written as
$3E_g=\Delta+\sqrt{\Delta^2+8J^2}=2\sqrt 2 J (z+\sqrt{z^2+1})$ 
or equivalently,
\be
   \frac{E_g}{2\sqrt 2 J}=\frac{e^v}{3}.
   \label{xxzgap.eqn}
\ee
The difference equation describing the flow of $v$ is,
\be
   v_{n+1}-v_n=\asinh\left[\frac{1}{2}\sinh v_n~~e^{2v_n}\right]-v_n,
\ee
which has been plotted in Fig.~\ref{xfixed.fig}.
The above equation has a repulsive fixed point at $v^*=\ln \sqrt2$ which corresponds to the
BKT point $z^*=1/(2\sqrt 2)$ or $\Delta=J$.

\begin{figure}[t]
\includegraphics[width=0.49\textwidth]{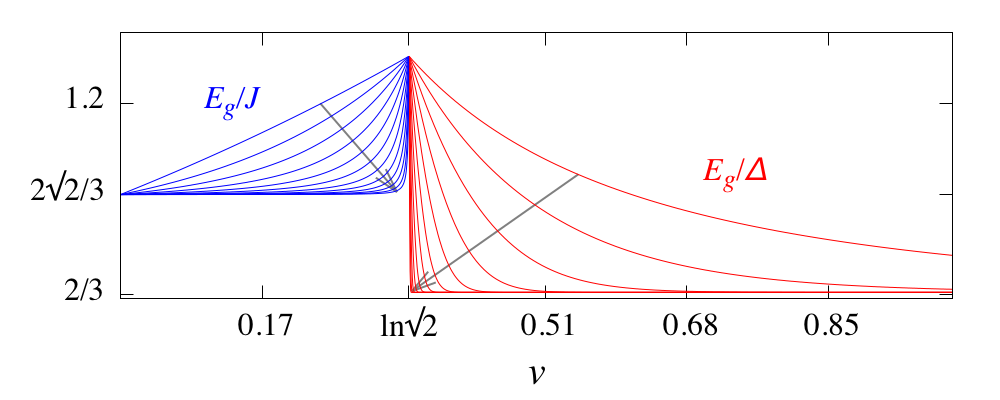}
\caption{(Color online) The evolution of gap function on two sides of the 
phase transition. The BKT point $J=\Delta$ corresponds to $v^*=\ln\sqrt2$. 
For clarity at every length scale the gap is normalized to the exchange $J$ (blue)
and the $\Delta$ (red) of the same length scale for the liquid and Mott insulating
sides, respectively. The gray arrows indicate the direction of RG flow. Note 
that in the liquid (blue) side the $J$ itself very quickly approaches zero at
large length scales, and hence the blue side is gapless. 
}
\label{xxzgap.fig}
\end{figure}

We have numerically evaluated the flow equation~\eqref{vflow.eqn} and have used it
to generate a flow for the gap of the XXZ model, Eq.~\eqref{xxzgap.eqn} in
Fig.~\ref{xxzgap.fig}. The flow of the energy gap for the left (right) of BKT point $v^*=\ln\sqrt 2$
has been plotted in blue (red). For clarity of presentation we have normalized the blue 
plots to $J$, while the red plots are normalized to $\Delta$ which is the dominant and 
natural energy scale on the right side of the BKT point. The direction of the flow has been
indicated by gray arrows. To the left of the BKT point $v^*=\ln\sqrt 2$, every thing flows to 
$v=0$ (corresponding to $z\sim \Delta/J=0$) and hence the relevant energy scale is $J$
which sets the scale of the energy gap. But on the other hand since for every $v$ one always
has $\cosh^2v>1$, the second of Eq.~\eqref{vflow.eqn} indicates that $J$ flows faster than the
geometric progression $J_{n+1}=J_n/2$ which implies that the energy scale $J$ in the large $n$ 
(long length) limit approaches to zero and therefore the left of the BKT point (blue) lines is 
a gapless phase. Indeed one can do a formal expansion 
around the non-interacting attractive fixed point at $v^*=0$ which gives,
\begin{align}
   &v_{n+1}=\left(\frac{1}{2}\right) v_n \Rightarrow v_n=\left(\frac{1}{2}\right)^n v_0\Rightarrow\nn\\
   & E_g=\frac{2\sqrt2}{3} e^{v_n}=\frac{2\sqrt2}{3}\left(e^{v_0\epsilon}\right),~~~~~\epsilon=\exp(-n\ln2)
\end{align}
where for large $n$, the quantity $\epsilon$ is exponentially small at large length scales. 
With this the behavior of gap function for the left of BKT point can be understood
and it can be seen why all the blue curves settle on the $E_g/J=2\sqrt2/3$ line. 

Now let us move to the right of BKT point in Fig.~\ref{xxzgap.fig} that corresponds to 
the Mott phase. Expanding around the BKT repulsive fixed point $v^*=\ln\sqrt 2$, we can write,
\begin{align}
   e^{v_n}&=e^{v^*}\exp\left[\kappa^n(v_0-v^*)\right],\label{vvs.eqn}\\
   J_n&=\exp\left[-3\kappa^n(v_0-v^*)\right],~~~n\gg 1\label{jjs.eqn}\\
   \frac{E_g}{\Delta}&=\frac{e^{v_n}}{3\sinh(v_n)}\label{ees.eqn},
\end{align}
where $\kappa=5/3$ and we have used $e^{2v^*}=2$ along with the fact that for large 
$n$, $\kappa^n-1\approx \kappa^n$. The first of the above equation shows that in the 
large $v$ limit where $\Delta/J\sim z=\sinh v\sim e^v/2$ diverges, the relevant energy 
scale is $\Delta$. The second equation indicates how does the scale $J$ fade away at
large length scales, and the third equation shows why in the $n\to\infty$ where $v\to\infty$,
the ratio of energy gap per site $E_g$ and $\Delta$ approaches the $2/3$ in agreement
with Fig.~\ref{xxzgap.fig}. However since the energy scale $\Delta$ in the right of 
BKT point is not renormalized to zero, the right of BKT point $v^*=\ln\sqrt2$ is 
actually a gapped phase corresponding to the Mott insulating phase of the underlying JW fermions.

Fig.~\ref{xxzgap.fig} nicely shows that for every $0<v<\ln\sqrt2$ corresponding to 
$0<\Delta<J$ the gap settles on $2\sqrt 2 J/3$ which eventually approaches to
zero as does $J$ in the long length limit. The closer $\Delta$ is to zero, earlier in
RG steps the gap reaches the zero. For values of $\Delta<J$ that are closer to $J$, 
a larger RG iteration, i.e. a longer length is required to attain the zero gap. But eventually
for long enough length scales, all Hamiltonians with $0<\Delta<J$ end up in a gapless state.
This gapless state terminates at the BKT point $\Delta=J$.

A further hallmark of the BKT transition is the behavior of the gap to the right of BKT 
point. In the MI phase the gap is essentially given by the $2\Delta/3$ at every length 
scale. The closer the $\Delta$ is to the right of BKT, it takes a longer length to
settle the gap to $2\Delta/3$ asymptote. Combining the asymptotic behavior of 
Eqs.\eqref{vvs.eqn},~\eqref{jjs.eqn},~\eqref{ees.eqn} we obtain
\begin{align}
   \ln E_g\sim &\ln\Delta=-3\ell^{1/\nu} (v_0-v^*),\nn\\
   &=-3\left(\frac{\ell}{\xi(v_0)}\right)^{1/\nu},\\
   &\nu=\ln3/\ln\kappa=2.150\nn
\end{align}
where $\nu$ is the correlation length exponent $\xi(v)\propto|v-v^*|^{-\nu}$ which
satisfies $\xi(v_{n+1})=\xi(v_n)/3$. This algebraic behavior of the logarithm of 
the gap is the three-site block-spin RG version of the BKT behavior~\cite{denNijs,Fradkin},
\be
   E_g(v)=\exp\left(-\frac{\rm const}{\sqrt{v_0-v^*}}\right).
\ee

Equipped with block-spin RG interpretation of the phase transitions of XY and XXZ models,
we are now prepared to construct a global phase diagram of the XYZ model. 

\section{Phase diagram of the XYZ spin chain}
\begin{figure}[t]
\includegraphics[width=0.50\textwidth]{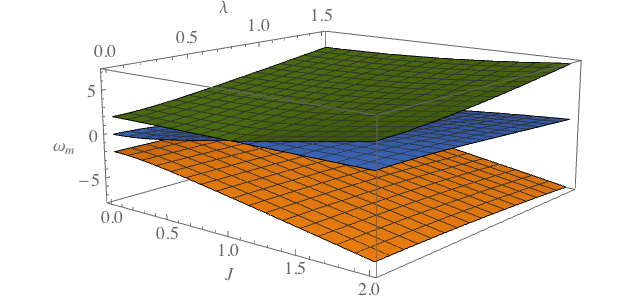}
\caption{(Color online) The evolution of roots as functions of $J$ and $\lambda$.
All three axes in this figure are in units of $\Delta$. The lowest root (orange)
correspond to $m=0$ and the highest root (green) correspond to $m=2$. 
}
\label{roots.fig}
\end{figure}

The role of $\lambda\ne 0$ when $\Delta=0$ is to open up a topologically non-trivial bulk-gap.
The role of $\Delta\ne 0$ when $\lambda=0$ on the other hands is to open up a an Ising gap, or
in the language of the massive Thirring model to open up a Mott gap. To distinguish this
Ising limit from the Ising limit of the XY model, we call the former Mott-Ising gap,
while for the gapped state due to p-wave pairing of JW fermions we use the term KI gap. 
Now it is desirable
to have both these gap opening mechanisms together, and to study the critical (gapless)
line separating these regions. For this purpose we need to analytically diagonalize the three-site 
Hamiltonian~\eqref{3site.eqn} which gives the following equation for the eigenvalues $\omega$:
\bearr
   &\omega^3-4(\Delta^2+2J^2+2\lambda^2)\omega-16\Delta(\lambda^2-J^2)=0\label{eigenval.eqn}.
\eearr
This is already in its canonical form $\omega^3-12P^2\omega+16Q=0$, which admits three 
trigonometric solutions,
\be
  \omega_m=- 4P\cos\left[\frac{1}{3}\arccos\left(\frac{Q}{P^3}\right) -\frac{2\pi m}{3}\right],
  \label{trig.eqn}
\ee
where $m=0,1,2$.
In the MI limit where hybridizations $\lambda$ and $J$ vanish, the roots of the cubic equation 
given by trigonometric formula~\eqref{trig.eqn} reduce to,
\be
   \omega_0\to -2\Delta,~~~~\omega_1\to 0,~~~~\omega_2\to 2\Delta.
\ee
In the MI limit the ground state is the root that starts at $-2\Delta$ and always remains
the ground state, except for the special point $J=0,\lambda=\Delta$ where the first excited
state touches the ground state. This is shown in Fig.~\ref{roots.fig}.
This indicates that $\omega_0$ always remains the ground state.
Therefore the ground state wave function corresponding to this energy is,
\begin{align}
   &\ket{\phi_+}=a\ket{\up\up\up}+b\ket{\dn\up\dn}+c(\ket{\up\dn\dn}+\ket{\dn\dn\up})/\sqrt2,\\
   &\ket{\phi_-}=a\ket{\dn\dn\dn}+b\ket{\up\dn\up}+c(\ket{\dn\up\up}+\ket{\up\up\dn})/\sqrt2,\\
   &a=2\sqrt2\lambda(\omega_0+2\Delta)/d^2,\nn\\
   &b=2\sqrt2J(\omega_0-2\Delta)/d^2,~
   c=(\omega_0^2-4\Delta^2)/d^2,\nn\\
   &d^2=\sqrt{8\lambda^2(\omega_0+2\Delta)^2+8J^2(\omega_0-2\Delta)^2+(\omega_0^2-4\Delta^2)^2},\nn
\end{align}
where the naming $d^2$ is chosen such that $d$ will have the dimension of energy.
Note that the above ground state is non-degenerate as long as $(J,\lambda)\ne(0,\Delta)$.
On the boundary of each cluster only the spins $\vec\sigma_0$ and $\vec\sigma_2$ are living which
might be connected to neighboring blocks. The symmetry under exchange of site indices $0,2$ in 
the cluster is manifest in the above Kramers doublet ground states. So we only need to compute the
transformation of one of them, e.g. $\vec\sigma_0$. The computation is straightforward noting
that every operation $\sigma^{x,y}$ changes the conserved quantity $\zeta$ and hence only has
off-diagonal components between the two ground states with $\zeta=\pm1$. For the same reason,
operator $\sigma^z$ does not change the charge $\zeta$ (fermion parity in the language of JW fermions)
and hence only has diagonal components. This gives,
\begin{align}
   &\sigma_0^{x} \to \sqrt 2 c (a+b)\sigma'^x,\nn\\
   &\sigma_0^{y} \to \sqrt 2 c (a-b)\sigma'^y,\nn\\
   &\sigma_0^z \to (a^2-b^2) \sigma'^z,\nn
\end{align}
which result in the flow equations,
\bearr
   &&J'+\lambda'=2(J+\lambda)c^2(a+b)^2,\\
   &&J'-\lambda'=2(J-\lambda)c^2(a-b)^2,\\
   &&\Delta'=\Delta(a^2-b^2)^2.
\eearr

To proceed further, let us define the dimensionless version of $\Delta$ and $\lambda$ in units of $J$ 
with $\Delta=xJ,\lambda=yJ$ . Then the dimensionless version of the flow equations become,
\begin{align}
   p&=\sqrt{\frac{x^2+2+2y^2}{3}},~~~q=x(1-y^2),\nn\\
   \eps&=-4p\cos\left[\frac{1}{3}\arccos\frac{q}{p^3}\right],\nn\\
   \eta&=\sqrt{8y^2(\eps+2x)^2+8(\eps-2x)^2+(\eps^2-4x^2)^2},\nn\\
   \alpha&=2\sqrt2 y (\eps+2x)/\eta,\nn\\
   \beta&=2\sqrt2(\eps-2x)/\eta,\nn\\
   \gamma&=(\eps^2-4x^2)/\eta,\nn\\
   x'&=\frac{x(\alpha^2-\beta^2)^2}{2\gamma^2\left[(\alpha^2+\beta^2)+2\alpha\beta y\right]},\\
   y'&=\frac{2\alpha\beta+y(\alpha^2+\beta^2)}{\alpha^2+\beta^2+2\alpha\beta y}.
\end{align}
where $\eps$ is the dimensionless ground state energy defined by $\eps=\omega_0/ J$.
Let us check the limit, $x\to 0$ of XY model where we obtain
$\eps\to-2\sqrt 2\sqrt{1+y^2}$ which implies $\alpha\to y/\sqrt{2(1+y^2)}$,
$\beta\to 1/\sqrt{2(1+y^2)}$, $\gamma\to -1/\sqrt 2$ whereby 
the flow equations give $x'=0$, $y'=y(y^2+3)/(1+3y^2)$ which is the same as Eq.~\eqref{recy.eqn}. 

\begin{figure}[t]
\includegraphics[width=.98\columnwidth]{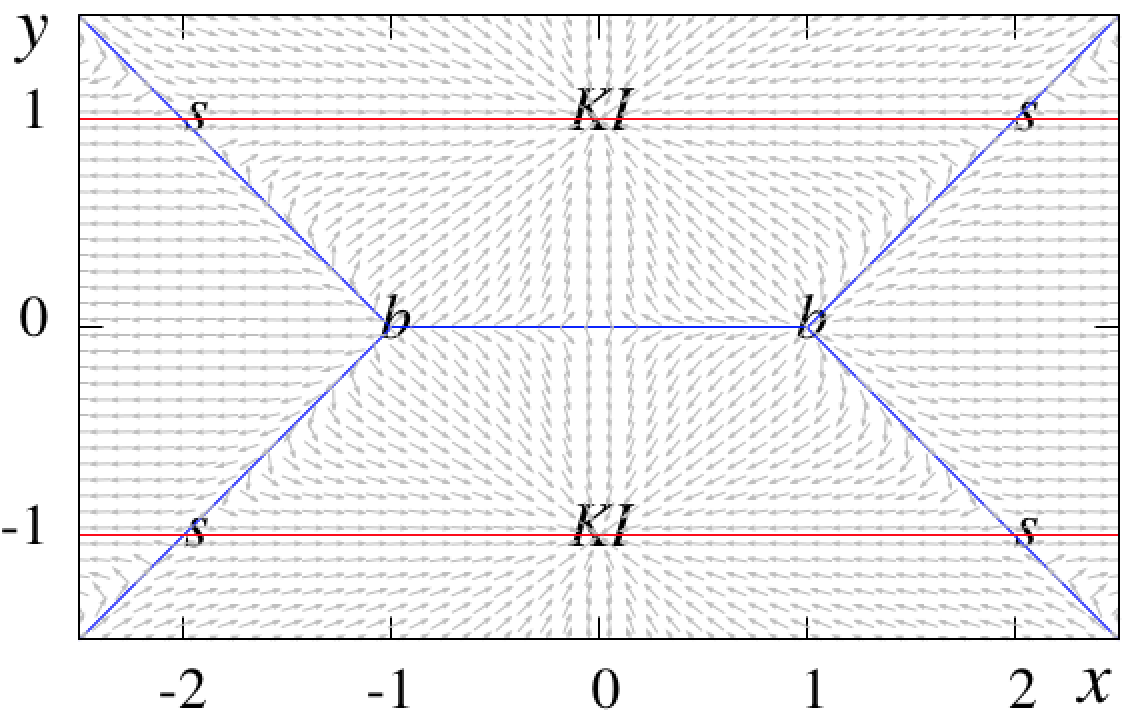}
\caption{(Color online) The phase portrait for the XYZ model in one dimension. 
The horizontal and vertical axes are $x=\Delta/J$ and $y=\lambda/J$, respectively.
The flow profile is symmetric with respect to the $y$ axis. The two attractive
fixed points at $(0,\pm 1)$ are Kitaev-Ising fixed points. The fixed points at
$x\to \pm \infty$ are Mott-Ising fixed points. The repulsive fixed point at 
$(1,0)$ are born Kosterlitz-Thouless transition of the XXZ model which are now
turned into tricritical points.
}
\label{flow.fig}
\end{figure}

The phase portrait of the above set of flow equations is shown in Fig.~\ref{flow.fig} where the 
horizontal axis denotes $x=\Delta/J$ and the vertical axis represents
$y=\lambda/J$. In the language of Jordan-Wigner fermions,
the $x$ is a measure of many-body interaction between the JW fermions, and $y$ controls the p-wave superconducting
interaction between these spin-less fermions. 
We colloquially use the term IO 
to refer to staggered magnetization in the case of AFM Ising point, and to magnetization
in the case of FM Ising point. Since these two are related by a canonical transformation,
we do not distinguish them and only refer to them by Ising order. 
For definiteness let us assume that $J>0$ and spell out the phase diagram
obtained by the block-spin RG method.

The phase portrait of the XYZ model in Fig.~\ref{flow.fig} is characterized by
Ising attractors at $(x,y)=(0,\pm 1)$  which are denoted as KI and $(x,y)=(\pm\infty,0)$ 
-- not shown in the figure -- and repellers
at six BKT points two of which are denoted by letter $b$ and the other four 
are along the asymptotes $|y|=|x|-1$. Therefore there are totally six BKT points
that when joined together determine the phase boundary depicted as blue line in
this figure. The BKT points at $(x,y)=(\pm 1,0)$ are tricritical points~\cite{Ercolessi2013}.

This blue line coincides with the exact phase boundary extracted
from the solution of Baxter~\cite{denNijs}.
There are four saddle points at $(x,y)=(\pm2,\pm1)$ that guide the 
flow lines. The blue gapless lines divide the plane of $x=\Delta/J$ and $y=\lambda/J$ 
into four regions. The Ising attractors at far right (left) of the $x=\Delta/J$ 
axis correspond to AFM (FM) Ising order. For $J>0$ ($J<0$), the KI attractors
at $(x,y)=(0,\pm 1)$ correspond to AFM (FM) Ising order along $\hat x$ and $\hat y$
directions. As we saw in section~\ref{XY.sec} the KI points are characterized with
a winding number corresponding to which a pair of MFs are spawn at the two ends
of the spin chain. Now the KI attractors with their non-trivial topology have
turned into global attractors in the plane of $\Delta$ and $\lambda$. 

The equation of the gapless phase boundaries (blue lines) of the phase portrait in 
Fig.~\ref{flow.fig} are given by,
\begin{align}
   |y|=\Theta(||x|-1|)
   \label{boundary.eqn}
\end{align}
which agrees with the exact solution~\cite{denNijs,Ercolessi2011} and should be compared
e.g. with Fig. 3 of den Nijs~\cite{denNijs}. 
The whole region $y>\Theta(|x|-1)$ is attracted to the KI point at $(x,y)=(0,+1)$ with
winding number $n_w=+1$ and IO along $\hat x$ direction.
The entire region $y<-\Theta(|x|-1)$ is attracted to the other KI point at $(x,y)=(0,-1)$
with winding number $n_w=-1$ and IO along $\hat y$ direction.
The rest of the plane for $|x|>|y|+1$ is attracted to the MI fixed point with winding 
number $n_w=0$ and IO along $\hat z$ direction. 
In the language of JW fermions it means that when a p-wave superconducting bulk gap
is opened by a non-zero $\lambda$, turning on the interaction $\Delta$ between the
JW fermions is not capable to close the gap and change the topological charge from
$n_w=\pm 1$ of the KI point to the $n_w=0$ of the MI phase, unless the interaction 
$\Delta$ is larger enough to satisfy $|\Delta|>|J|+|\lambda|$.

The repulsive fixed pint at $(\Delta=J,\lambda=0)$ corresponds to the
BKT transition from critical (gapless) phase to the massive MI phase. The field theory
treatment in terms of sine-Gordon theory gives a critical value $\Delta_c=\pi J/2$~\cite{Fradkin}, 
while the exact solution gives $\Delta_c=J$~\cite{BaxterXYZ}. As discussed in previous sections, 
the fixed points 
at $(\Delta=0,\lambda=\pm J)$ correspond to Ising fixed points of the Kitaev chain which have now
turned into globally attractive fixed points. These points are gapped Ising phases, however to distinguish
them from the Mott-Ising phase, it is appropriate to call them Kitaev-Ising points. 
The present picture means that the KI fixed points
obtained from the XY limit that corresponds to Ising magnets polarized in $\hat x$ or $\hat y$ directions, 
and is entitled to a non-zero winding number, remain attractive fixed points in a broader parameter
range where the interaction between JW fermions ($\Delta$) can also 
be present. For interacting XYZ chain the $\hat x$ or $\hat y$ polarized KI fixed points 
remain the flow destination as long as the pairing interaction $\lambda\sim y$ is strong enough to 
satisfy $|y|>|x|-1$. Otherwise the Mott-Ising fixed point that polarizes the system along 
the $\hat z$ axis will win.

\section{Discussions and summary}
We have analyzed the phase diagram of the XYZ model. In the limiting case of the XY 
spin chain that corresponds the Kitaev chain model of spinless JW fermions paired
with p-wave superconducting interaction $\lambda$, we find that the Ising limit of 
the Kitaev chain that leaves a pair of sharply localized Majorana fermions in the 
two ends of the chain is actually the RG fixed point. In the XY limit we were further
able to analytically solve the flow equation. This allowed us to identify a geometric
progression inherent in the RG flow as a length scale associated with zero modes of
the system, namely the size of Majorana fermions. The Kitaev-Ising fixed points of 
the XY limit are characterized by a non-zero winding number. We further find that 
within the three-site cluster employed in our analysis the -- superconducting -- gap
at non-zero values of $\lambda$ develops as $\lambda^{\ln 2/\ln 3}$. 

The other extreme limit is that of the XXZ chain where the exact BKT point at
$\Delta=J$ is obtained from a block spin RG based on the three site problem. 
We analytically obtain the asymptotic behavior of the RG flow which enables us to 
establish the $\Delta>J$ region is gapped and flows to the Mott-Ising fixed point. 
Then we considered the effect of nonzero $\lambda$ and $\Delta$ which in the language
of JW fermions corresponds to the massive Thirring model. In this general case we find that
the ground state of the XYZ model is essentially gapped. The phase portrait is 
characterized by Ising attractors and BKT repellers. The gapless (blue) lines in 
Fig.~\ref{flow.fig} is essentially the exact result of Baxter. But the new insight
of the present analysis is that our phase portrait attaches topological significance
to the Baxter's exact solution. Indeed in the $y>\Theta(|x|-1)$ the KI fixed point
with winding number $n_w=+1$ which 
corresponds to Ising order along $\hat x$ direction is a global attractor,
while in the $-y>\Theta(|x|-1)$ region the KI fixed point with winding number $n_w=-1$
is a global attractor in the space of parameters $\Delta,\lambda$. For very strong
$|\Delta|$, the Mott-Ising attractor takes over which in characterized with a 
winding number $n_w=0$. Therefore the blue lines in Fig.~\ref{flow.fig} divide the
parameters space of the XYZ model into regions where across the (blue) border 
a winding number changes, and hence the transition from one gapped (Ising ordered)
state to another gapped state is actually a topological phase transition. 
The underlying topology explains why a simple three-site problem is able to capture the {\em exact}
phase diagram of the model.

The KI fixed points spawn a pair of Majorana fermions sharply
localized in the chain ends. Going from the KI fixed point e.g. at $(x,y)=(0,+1)$
to the other one corresponds to changing the polarization direction from $\hat x$ to
$\hat y$ direction. This in the language of Majorana fermions corresponds to exchanging
the two Majorana fermions of type $a$ and $b$ in the opposite ends of the chain
which requires the change of topology. 

The picture presented so far relies on the $\sz$ basis used in our analysis. 
Indeed thinking in terms of the couplings $J_x=J+\lambda$, $J_y=J-\lambda$, $J_z=\Delta$,
the three Ising limits can be mapped to each other by coordinate transformation.
Therefore assigning the three winding numbers $n_w=\pm 1$ to IO along $\hat x$
and $\hat y$ and $n_w=0$ to IO along $\hat z$ is a matter of choice. The reason is 
that the JW fermions and their associated MFs are constructed from the transverse
spin variables $\sx,\sy$. To that extent even the Mott-Ising phase at large $\Delta$
can be thought of a KI point when expressed in terms of JW fermions constructed 
from, e.g. $\sz,\sx$ variables. {\em Hence the Mott phase of the Thirring model
is entitled to have zero modes and hence is topologically non-trivial}. 
Indeed in one dimensional helical liquids a topologically non-trivial 
gap can be opened by two-particle interactions~\cite{Sela}.
To see how the above symmetry with respect to choice of the coordinate system
is reflected in the phase diagram of Fig.~\ref{flow.fig}, let us consider a portion
of the blue phase boundary that connects the two BKT points marked as b in the figure. 
Equation of this phase is $J_x=J_y,  |J_z|<|J_x|$. Changing the coordinate system
the equation of the boundary line would be $J_{x,y}=J_z$, which means $J\pm\lambda=\Delta$,
which then gives $x\mp y=1$ that is nothing but the equation of the portion of the blue line
emanating from the BKT point $(x,y)=(1,0)$ to the up-right and down-right of the figure.

It has been recently found that the ground state of the XYZ chain has non-trivial multi-fractality 
spectrum~\cite{Atas1,Atas2} which is entirely different from the type of multi-fractal behavior
in (disordering) Anderson transition and might have to do with the many-body localization~\cite{Fisher,Slagle}.
It is therefore desirable to develop an understanding of the XYZ model from the perspective
of topology which can also shed light on the role of topology in the corresponding problem of
interacting fermions or bosons. 
It would be interesting to examine the role of topology -- that can be diagnosed
by the bipartite charge fluctuation~\cite{Slagle} -- in the multi-fractal
behavior of the ground state. 

To summarize, we have obtained the {\em exact} phase diagram of the XYZ spin
chain. The gapless lines correspond to topological phase transitions through which
the appropriate Majorana zero modes are exchanges across the chain ends. 

\section{acknowledgements}
I thank Takanori Sugimoto, B. Normand and T. Farajollahpour for discussions. 
This work supported by Alexander von Humboldt fellowship for experienced researchers.


\begin{thebibliography}{50}
\bibitem{BaxterPRL} R. J. Baxter, Phys. Rev. Lett. (1971) {\bf 26}, 832.
\bibitem{Baxter8V} R. J. Baxter, Ann. Phys. (1972) {\bf 70}, 193.
\bibitem{BaxterXYZ} R. J. Baxter, Ann. Phys. (1972) {\bf 80}, 323.
\bibitem{McCoy} J. Johnson, S. Kriksky, and B. McCoy, Phys. Rev. A {\bf 8}, 2526 (1973).
\bibitem{ODBA} Y. Wang, W. -L. Yang, J. Cao, K. Shi, {\em Off-Diagonal Bethe Ansatz for Exactly Solvable Models}
Springer, 2015
\bibitem{odba-npb} J. Cao, W. -L. Yang, K. Shi, Y. Wang, Nucl. Phys. B. (2013) {\bf 877} 152. 
\bibitem{LSM} E. Lieb, T. Schultz and D. Mattis, Ann. Phys. (1961) {\bf 16} 407.
\bibitem{Mattis2006} D. C. Mattis, {\em The theory of magnetism made simple}, Word Scientific, 2006.
\bibitem{Raghu2011} Yuezhen Niu, S. B. Chung, Chen-Hsuan Hsu, I. Mandal, S. Raghu and S. Chakravarty,
   Phys. Rev. B {\bf 85} (2012) 035110.
\bibitem{NXY} S. A. Jafari, F. Shahbazi, Sci. Rep. {\bf 6}, 32720 (2016).
\bibitem{Sen} W. DeGottardi, M. Thakurathi, S. Vishveshwara, D. Sen, Phys. Rev. B {\bf 88} 165111 (2013).
\bibitem{Kitaev2001} A. Yu. Kitaev, Phys. -Usp. (2001) {\bf 44} 131.
\bibitem{Fradkin} E. Fradkin, {\em Field theories of condensed matter physics}, Cambridge University Press, 2nd Ed., 2013.
\bibitem{Luther} A. Luther, Phys. Rev. B (1976) {\bf 14} 2153.
\bibitem{Bergknoff} H. Bergknoff, H. B. Thacker, Phys. Rev. D (1979) {\bf 19} 3666.
\bibitem{Tsvelik} A. M. Tsvelik, {\em Quantum Field Theory in Condensed Matter Physics}
Cambridge University Press, 2003
\bibitem{Gogolin} A. O. Gogolin, A. A. Nersesyan, A. M. Tsvelik, {\em Bosonization and Strongly Correlated Systems}
Cambdige University Press, 1998.
\bibitem{denNijs} M. P. M. den Nijs, Phys. Rev. B (1981) {\bf 23} 6111.
\bibitem{Rigol} M. Dalmonte, J. Carrasquilla, L. Taddia, E. Ercolessi, M. Rigol,
Phys. Rev. B (2015) {\bf 91} 165136.
\bibitem{Langari2009} M. Kargarian, R. Jafari, A. Langari, Phys. Rev. A, (2009) {\bf 79} 042319.
\bibitem{Langari98} A. Langari, Phys. Rev. B (1998) {\bf 58} 14467.
\bibitem{Song2013} Xue-ke Song, T. Wu, L. Ye, Eur. Phys. J. D (2013) {\bf 67} 96.
\bibitem{Ercolessi2011} E. Ercolessi, S. Evangelisti, F. Franchini, F. Ravanini, Phys. Rev. B (2011) {\bf 83} 012402.
\bibitem{Ercolessi2013} E. Ercolessi, S. Evangelisti, F. Franchini, F. Ravanini, Phys. Rev. B (2013) {\bf 88} 104418.
\bibitem{Langari2004} A. Langari, Phys. Rev. B (2004) {\bf 69} 100402(R).
\bibitem{Alicea} J. Alicea, Rep. Prog. Phys. (2012) {\bf 75} 076501.
\bibitem{Strogatz} S. H. Strogatz, {\em Nonlinear Dynamics and Chaos}, Perseus Books, 1994.
\bibitem{Akbari} R. Jafari, A. Langari, A. Akbari, Ki-Seok Kim, J. Phys. Soc. Jpn. {\bf 86} (2017) 0204008.
\bibitem{Sela} E. Sela, A. Altland, R. Rosch, Phys. Rev. B {\bf 84} (2011) 085114.
\bibitem{Atas1} Y. Y. Atas, E. Bogomolny, Phys. Rev. E (2012) {\bf 86} 021104.
\bibitem{Atas2} Y. Y. Atas, E. Bogomolny, Phil. Trans. R. Soc. A (2014) {\bf 372} 20120520.
\bibitem{Fisher} J. R. Garrison, R. V. Mishmash, M. P. A. Fisher, Phys. Rev. B {\bf 95} (2017) 054204.
\bibitem{Slagle} K. Slagle, Y.-Z. You and C. Xu, Phys. Rev. B {\bf 94} (2016) 014205.
\end{thebibliography}
\end{document}